\documentclass[twocolumn,aps,prl,groupedaddress,showpacs, superscriptaddress]{revtex4-1} 

\usepackage{amsmath}
\usepackage{amssymb}
\usepackage{txfonts}
\usepackage{pxfonts}
\usepackage{gensymb}
\usepackage{graphicx,bm,units,yfonts}
\usepackage[table]{xcolor}
\usepackage{hyperref}
\usepackage{movie15}

\begin{document}

\title{Experimental Demonstration of Dynamic Topological Pumping Across Incommensurate Bilayered Acoustic Metamaterials}

\author{Wenting Cheng}
\affiliation{Department of Physics, New Jersey Institute of Technology, Newark, NJ, USA}

\author{Emil Prodan}
\email{prodan@yu.edu}
\affiliation{Department of Physics, Yeshiva University, New York, NY, USA}

\author{Camelia Prodan}
\email{cprodan@njit.edu}
\affiliation{Department of Physics, New Jersey Institute of Technology, Newark, NJ, USA}

\begin{abstract}
A Thouless pump can be regarded as a dynamical version of the integer quantum Hall effect. In a finite-size configuration, such topological pump displays edge modes that emerge dynamically from one bulk-band and dive into the opposite bulk-band, an effect that can be reproduced with both quantum and classical systems.  
Here, we report the first un-assisted dynamic energy transfer across a metamaterial, via pumping of such topological edge modes. The system is a topological aperiodic acoustic crystal, with a phason that can be fast and periodically driven in adiabatic cycles. When one edge of the metamaterial is excited in a topological forbidden range of frequencies, a microphone placed at the other edge starts to pick up a signal as soon as the pumping process is set in motion. In contrast, the microphone picks no signal when the forbidden range of frequencies is non-topological.
\end{abstract}

\maketitle

\begin{figure*}[t]
\includegraphics[width=\textwidth]{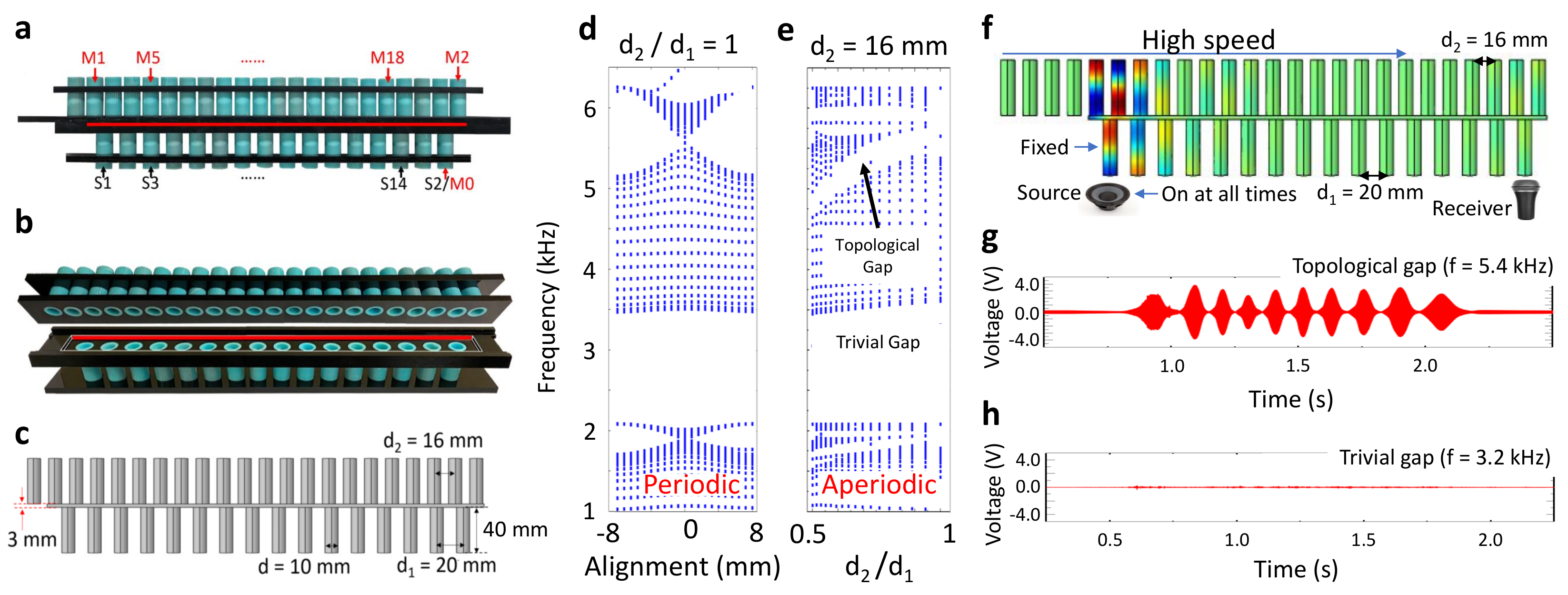}
\caption{\small {\bf Dynamic topological pumping.} (a) Photograph of a fully assembled acoustic bilayer consisting of top/bottom periodic arrays of cylindrical acoustic resonators with incommensurate lattice constants. The labels S$i$ and M$i$ indicate the positioning of the speaker and microphone during various experiments referenced in the text. The middle red bar indicates the presence of an inner chamber, which connects the top and bottom resonators and is referred to as the spacer. For dynamical pumping, additional resonators are mounted on the top left side. (b) Photograph of the inner structure, with the spacer now fully visible. (c) Cross-section showing only the domain of wave propagation, together with relevant parameters. Note that the very left resonator is completed decoupled from the main structure. (d) Bulk resonant spectrum as function of top-bottom relative alignment, when the top and bottom lattice constants are equal. In this case, all spectral gaps are trivial. (e) Bulk resonant spectrum as function of $d_2/d_1$. In this case, additional gaps open in the spectrum, which are all topological. (f) Schematic of the pumping process as well as a simulation of the air pressure at the beginning of the pumping cycle. (g) Microphone reading when the source frequency is adjusted in a topological gap. (h) Microphone reading when the source frequency is adjusted in a non-topological gap.} 
\label{Fig1}
\end{figure*}

\maketitle

More than 35 years ago, Thouless asked himself what happens with a filled sea of fermions when the underlying potential is slowly and periodically modulated in time \cite{ThoulessPRB1983}. He predicted that a precise non-fluctuating number of particles will be effectively transported from one side of the system to the other and that this number is determined by a topological invariant computed for a virtual system of one dimension higher than the original. The effect has been directly demonstrated recently, with both fermions and bosons \cite{NakajimaNatPhys2016,LohseNatPhys2016}. It is now well established \cite{XiaoRMP2010, ProdanPRB2015} that augmentation of a parameter space to a $d$-dimensional quantum or classical system can give access to topological effects that, in normal conditions, are observed in $d+1$ or higher dimensions. The prototypical example is the periodic 1-dimensional Rice-Mele model \cite{RicePRL1982}, where an adiabatic deformation of the parameters leads to a virtual 2-dimensional system whose energy bands support non-trivial Chern numbers \cite{XiaoRMP2010}. As a result, the system displays chiral edge bands when driven in an adiabatic cycle and edge-to-edge topological pumping becomes possible. 

The existing experimental works on edge-to-edge topological pumping can be classified in three groups. The ones in the first group \cite{ApigoPRM2018,ApigoPRL2019,NiCommPhys2019,VossPLA2020,hafezi2013imaging,xia2020topological} report only renderings of the resonant spectra as functions of the adiabatic parameters. The parameters are not varied continuously but rather the measurements are interrupted and the systems are adjusted by hand or other means to achieve the next parameter values. Acquisition of the spectra for a single adiabatic cycle can take days. In the experimental works from the second group \cite{KLR2012,RuzzenePRL2019,RivaPRB2020,shen2019one,zilberberg2017photonic,lustig2019photonic,verbin2015topological}, the systems with different adiabatic parameters are rendered and coupled in space and the profiles of the resonant modes are mapped in space rather than time. The connection with a real dynamical Thouless pump is done through a mathematical argument that involves simplifications and assumptions \cite{KLR2012}. Lastly, the experimental works in the third group \cite{GrinbergNatComm2020,XiaArxiv2020} report assisted dynamical edge-to-edge pumping. We call it {\it assisted} primarily because energy was pumped into the mode to keep it alive as it traversed from one edge to another. Without such external intervention, the pumping would have succumbed to the dissipation and nothing would have been observed at the receiving end of the system. These experiments also contain a large number of active components controlled by an expensive layer of electronics, whose complexity grows with the size of the system. For this reason, the system in \cite{GrinbergNatComm2020} had only eight unit cells. While valuable demonstrations, these approaches do not offer yet a path towards practical implementations.

We demonstrate an un-assisted edge-to-edge topological pumping of sound, where the human intervention is completely absent once the mode is loaded at one end of the system. The key innovation is the use of an aperiodic meta-material structure that has a simple built-in mechanism that implements global structural changes resulting in rapid and repeated cyclings of its phason. This mechanism is simply the relative sliding of two coupled incommensurate periodic acoustic crystals. All the previous experimental works based on aperiodic structures employ the quasi-periodic pattern originally proposed in \cite{KLR2012}, where the dynamical matrices can be directly connected with the Aubry-Andr\'e and Harper models \cite{AubryAIPS1980,Harper1955}. This is not the case for an incommensurate bilayered acoustic crystal, yet we show in the Supplementary Materials that its dynamical matrix belongs to an algebra isomorphic to that of magnetic translations. As such, the spectral gaps carry Chern numbers and the bulk-boundary correspondence principle is the same as for the Chern insulators. However, our topological system is distinct from the standard Chern mechanical crystals \cite{ProdanPRL2009, WangPRL2015,NashPANS2015} because one dimension is virtual.

Using the phason of a quasi-periodic structure to generate topological edge modes is an appealing strategy because the bulk resonant spectrum is independent of the phason \cite{Footnote1}, hence {\it all} the bulk spectral gaps are preserved during the phason cycles. This remarkable characteristic is un-matched by any other design principle \cite{Footnote2}. Furthermore no fine-tuning of the systems is required \cite{ApigoPRM2018}, because the presence of topological edge modes steams from the aperiodic pattern alone.  In \cite{ProdanJGP2019}, we supplied an algorithmic method to generate aperiodic system with phasons living on generic topological spaces. In particular, \cite{ProdanJGP2019} idenfied the phason space for incommensurate bilayered patterns. The principles discovered in \cite{ApigoPRM2018} and \cite{ProdanJGP2019} made the present work possible.


\begin{figure*}[t]
\includegraphics[width=0.9\linewidth]{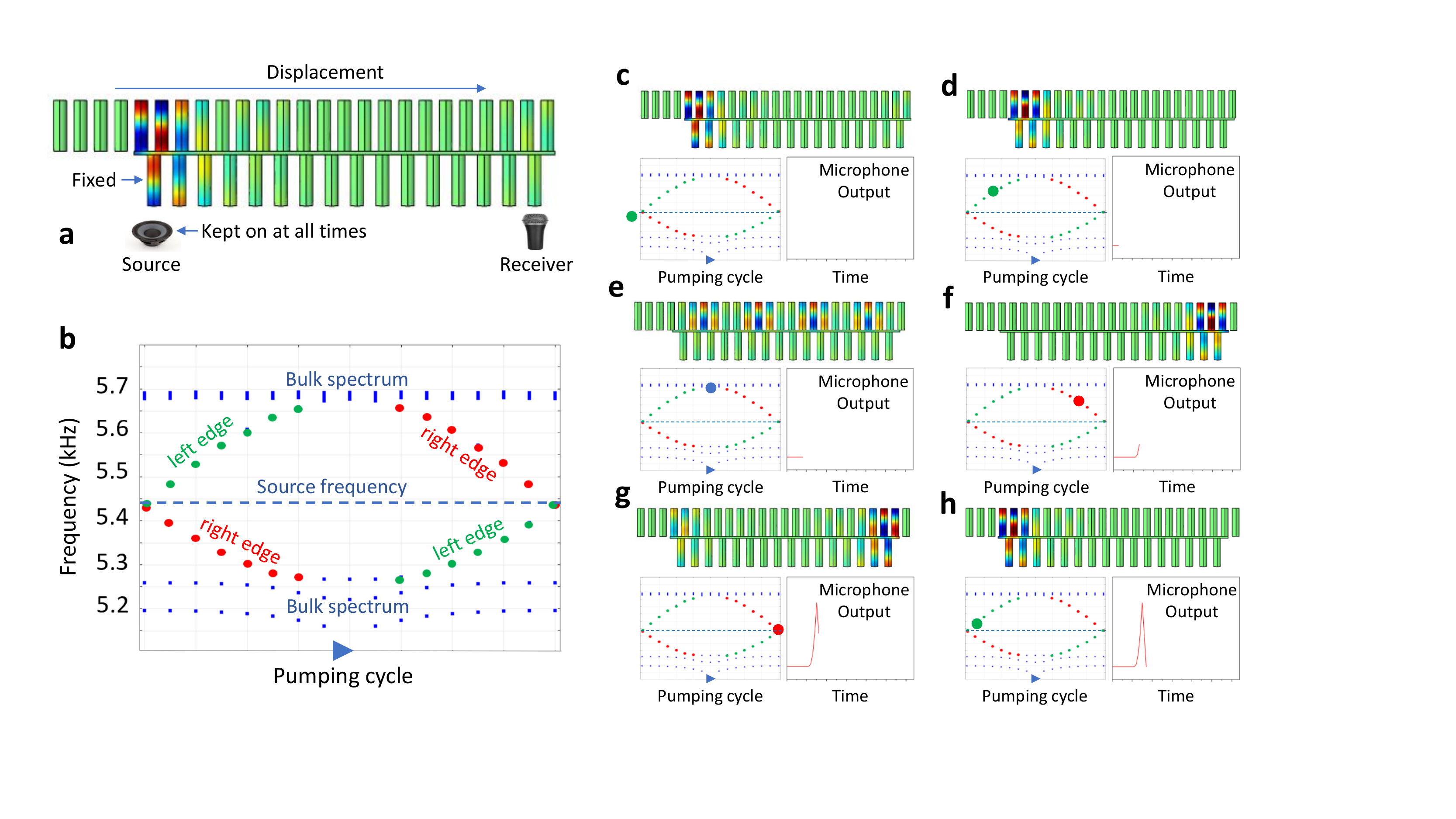}
\caption{\small {\bf Principles and mechanism of our dynamical pumping.} (a) The configuration of the system at the beginning of a pumping cycle. The top array is uniformly displaced to the right and, after a total displacement $d_2$, the system returns in its original configuration and completes a full pumping cycle. A speaker is inserted in resonator S1 and is kept on at all times, while a microphone is inserted in the resonator M0. (b) Simulation of the resonant spectrum as function of displacement. Chiral left and right edge bands are observed, which both connect two disjoint parts of the bulk spectrum. (c-h) Rendering of pumping mechanism: The left edge mode is loaded when the source frequency matches the mode frequency (c); The mode self-oscillates while its frequency is pushed up (d); The character of the mode changes from left-localized to delocalized (e); The character of the mode changes again from delocalized to right-localized (f); The mode self-oscillates as its frequency is pushed down (g); The cycle repeats itself as the top array is further pushed to the right (h).  The microphone starts to pick signal after the event (e). The simulations in panels (c-h) show the spatial profile of the resonant mode highlighted in the sub-panel below it. The shown microphone outputs are not from real measurements. The pumping parameter is $\varphi$ in all panels.} 
\label{Fig2}
\end{figure*}

Our main results are summarized in Fig.~\ref{Fig1}, where we present direct evidence of energy transfer from one end of a bulk structure to the other end, even though the frequency of the source falls in a forbidden wave-propagation range. This energy transfer happens in pumping conditions and when the source frequency is  in a topological spectral gap of the metamaterial. In contradistinction, when the frequency is adjusted in a non-topological spectral gap, there is no energy transfer even though same conditions of pumping are applied. Our experimental platform consists of the two incommensurate periodic arrays of acoustic resonators described in Fig.~\ref{Fig1}(a-c). The dimensions have been optimized to maximize the size of the topological gap. This type of patterned resonators was theoretically studied in \cite{ProdanJGP2019}, where it was found to support topological spectral gaps and topological edge modes. However, to our knowledge, this is the first time when coupled incommensurate chains are experimentally used to engineer chiral edge bands for topological pumping. 

Key to our experimental design was the replacement of any elaborate interconnections between acoustic resonators with a thin uniform spacer, extending from one end of the structure to the other. The resonators are attached to and coupled through this spacer. Note that this type of coupling does not allow fine-tuning but, as we mentioned, that is not necessary when using aperiodic principles, as long as the coupling is strong \cite{ApigoPRM2018}. Furthermore, edges can be created by simply filling the spacer with solid material. The adiabatic parameter of the system is $\varphi = x/d_2$, where $x$ and $d_2$ are specified in Fig.~\ref{Fig1}(c). Note that $\varphi$ lives on the circle, which is only a part of the total phason space \cite{Footnote3}. The advantages of our design are: a) $\varphi$ can be driven in an adiabatic cycle by simply sliding the top array while holding the bottom one fixed; b) Since the bottom array is fixed, we can continuously pump energy at one edge (and only on the that edge) by placing a source on the first bottom resonator; c) The left and right edges can be independently adjusted to achieve the optimal dispersion of the edge modes.

The numerically simulated topological pumping process is reported in Fig.~\ref{Fig2}, where we also explain its mechanism. Sure enough, the left and right chiral edge bands are present in the topological gap. Note their particular and optimal dispersion, which made the dynamical pumping possible. Indeed, it is important that the right chiral edge band emerges from the top bulk-band shortly after the left chiral edge dived into the same band. This is because the non-adiabatic effects cannot be prevented when the pumping of energy is through the bulk states. As such, one has to optimize the pumping cycle such that there is a rapid change of the mode character from left-localized to extended and to right-localized, exactly as it can be seen in Fig.~\ref{Fig2}(c-h), where our pumping cycle was broken down into steps. In a standard topological edge-to-edge pumping, the mode self-oscillates after being loaded at the left edge, hence the pumping cycle must be performed fast enough to overcome dissipation.

Given the particular engineering of our system, the pumping cycle can be performed extremely fast and repetitively, even without any external intervention. This enabled us to achieve the first un-assisted dynamical energy pumping via topological edge modes. Its dramatic manifestation is documented in Fig.~\ref{Fig1}(g), where a receiver placed opposite to an acoustic source is shown to pick up acoustic signal when the excitation frequency is in a topological resonant gap. In this experiment, 10 resonators were added beyond the edge to the left side of the top array, which resulted in the 10 pumping cycles visible in Fig.~\ref{Fig1}(g). The time  period of the pumping cycle is approximately 0.12 seconds in Fig.~\ref{Fig1}(g). We have experimented with the time period of the cycle and found that the energy transfer is completely cut out when the period is about 1 second. This demonstrates that the pumping process is indeed essential for the energy transfer across the acoustic meta-crystal. Furthermore, when the source frequency is adjusted in a non-topological spectral gap, the receiver picks no signal whatsoever. We have experimented with different source frequencies inside the non-topological gap and we can confirm that the receiver does not pick any signal even when the frequency is very close to the bulk spectrum. This demonstrates that the chiral edge bands, formed inside the topological spectral gap, play an essential role for the energy transfer phenomena detected in our experiments. 

The sound of the pumping reported in Fig.~\ref{Fig1}(h) can be played here, $
\includemovie[poster=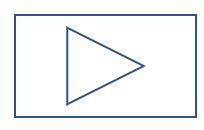,repeat=10]{0.03\linewidth}{0.03\linewidth}{Fig1h.wav}$ , or from the audio files available online. As one can see, there is stark difference between the two pumpings reported in Fig.~\ref{Fig1}. Taking into account all the above facts, there can be no doubt that the energy transfer across the meta-crystal was through a classic topological pumping process.

\begin{figure}[t]
\includegraphics[width=\linewidth]{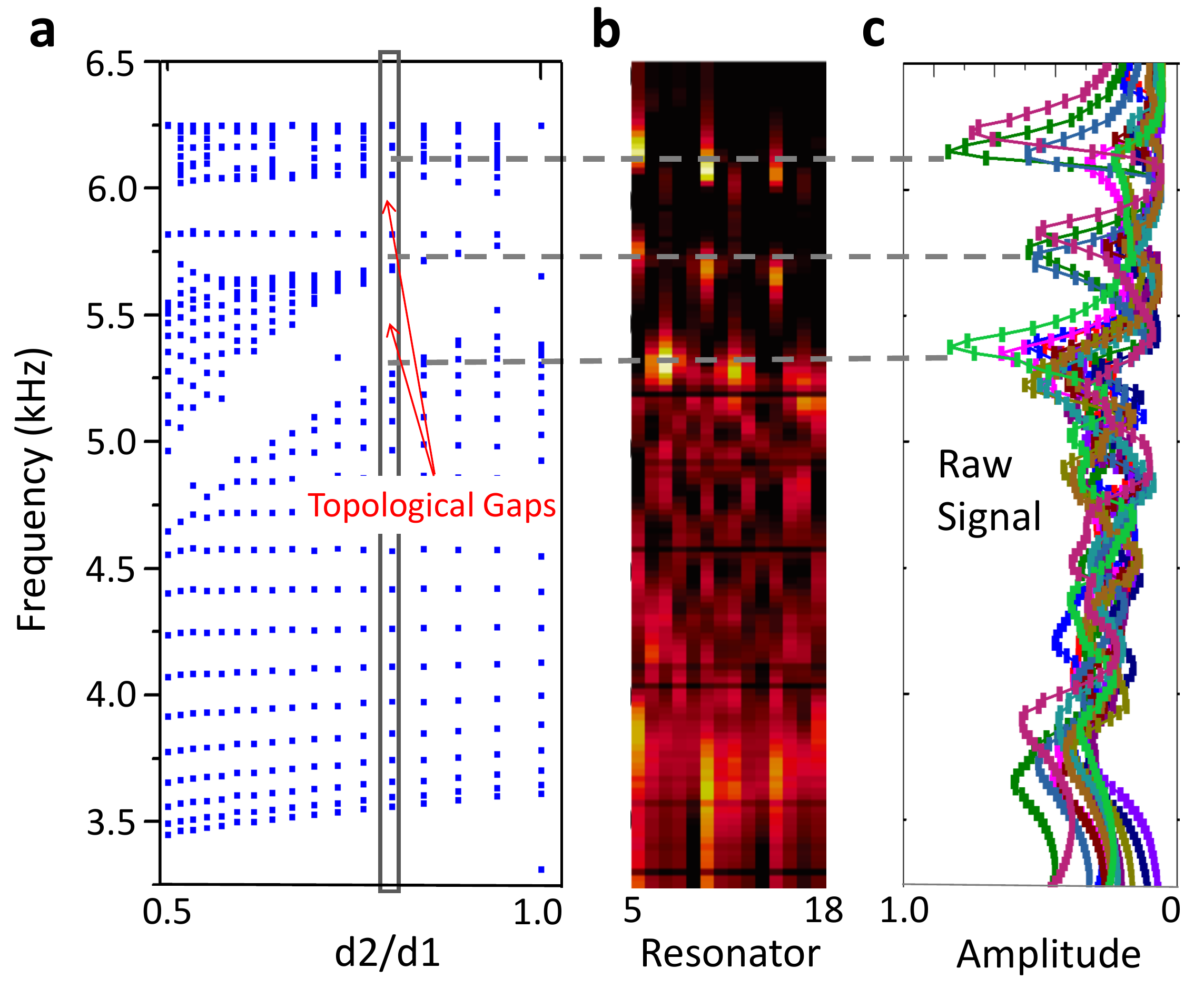}
\caption{\small {\bf Experimental mapping of the bulk resonant spectrum.} (a) Simulated resonant spectrum reproduced from Fig.~\ref{Fig1}(e), with arrows indicating the topological gaps. The vertical marking identifies $d_1 = 20$~mm and $d_2 = 16$~mm used in experiments. (b) Measurement of the spaced-resolved density of states (See Supplementary Material for Experimental Protocols). (c) Collapse of the data in panel (b) on the frequency axis. Two spectral gaps can be clearly identified in the experimental data, which are well aligned with the theoretical calculations.} 
\label{Fig3}
\end{figure}

As we already mentioned,  the resonator coupling through the spacer does not allow fine-tuning but that is not necessary when using aperiodic principles, as long as the coupling is strong \cite{ApigoPRM2018}. To understand the mechanism of topological gap generation in our system, we show first in Fig.~\ref{Fig1}(d) the evolution of the simulated resonant spectrum with respect to the relative alignment of two identical arrays of resonators. As expected in any 1-dimensional periodic system, gaps appear in the resonant spectrum and, as the system switches between period-one and period-two, some of these spectral gaps close while other remain open. Regardless of that behavior, all these gaps are topologically trivial because  the resonant bands seen in Fig.~\ref{Fig1}(d) result from dispersion-induced thickening of the discrete resonances of the individual resonators. However, when the lattice constant of the bottom array is varied and the system becomes aperiodic, these trivial bands are seen in Fig.~\ref{Fig1}(e) to become fragmented in sub-bands, exactly as it happens when a magnetic field is turned on a two dimensional electronic system \cite{HofstadterPRB1976}.  In the Supplementary Material, we in fact show that the dynamical matrix behind the resonant spectrum belongs to an algebra of observables generated by two operators obeying the same commutation relations as the magnetic translations. The conclusion is that the spectrum seen in Fig.~\ref{Fig1}(e) is a representation of the Hofstadter butterfly \cite{HofstadterPRB1976}. In particular, the sub-bands carry non-zero Chern numbers \cite{ApigoPRM2018,ProdanJGP2019} and the presence of the chiral edge bands can be explained by the standard bulk-boundary correspondence \cite{KellendonkRMP2002,ProdanSpringer2016}.

\begin{figure}[t]
\includegraphics[width=\linewidth]{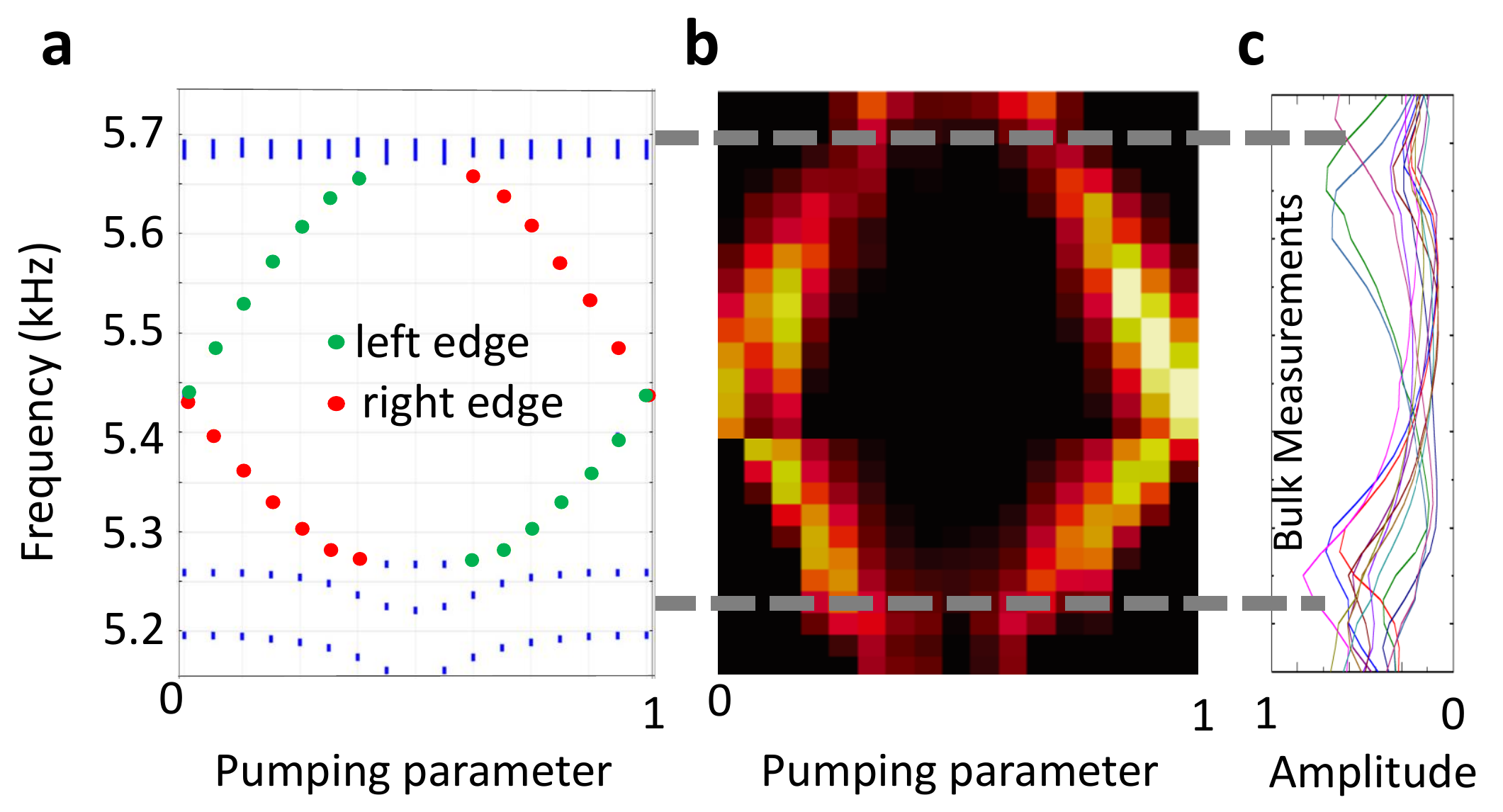}
\caption{\small {\bf Experimental measurement of the topological edge bands.} (a) Simulated resonant spectrum for the finite acoustic meta-crystal shown in Fig.~\ref{Fig1}(a), as function of the pumping parameter. The latter is the ratio between the displacement $ x$, shown in Fig.~\ref{Fig1}(c), and $d_2$. (b) Experimental measurement of the resonant spectrum as function of the pumping paramter. (c) Bulk-spectrum measurements reproduced from Fig.~\ref{Fig3}(c), used here to pin-point the position of the bulk bands shown by dashed lines.} 
\label{Fig4}
\end{figure}

The simulated bulk spectrum is reproduced with high fidelity by the experimental measurements, as demonstrated in Fig.~\ref{Fig3}. In particular, well defined bulk-spectral gaps can be identified in the measured local density of states, which are well aligned with the theoretical predictions. The frequency 5.4~kHz used for topological pumping in Fig.~\ref{Fig1}(g) falls in the middle of one such gap. Furthermore, the signature of the non-zero Chern numbers, that is, the chiral edge bands, are also detected experimentally, as reported in Fig.~\ref{Fig4}. By comparing the panels (a) and (b), one can see that the experiment reproduces the simulations with very high fidelity.

Having demonstrated an un-assisted energy transfer via a topological pumping process, we have laid down a set of specific principles which could facilitate the engineering of the effect in many other contexts. The most important one is that fine-tuning is not necessary which, together with the many different ways of engineering phason spaces \cite{ProdanJGP2019}, relaxes the design constraints, hence giving scientists better chances with finding optimal and practical meta-structures. While for meta-materials this process is now more or less straightforward, it will be extremely interesting if these aperiodic principles can be successfully applied to mesoscopic systems and achieve electron pumping in conventional insulators.

\vspace{0.2cm}  

\acknowledgments{All authors acknowledge support from the W. M. Keck Foundation. E. P. acknowledges additional support from the National Science Foundation through the grant DMR-1823800.}

\end{document}